\documentstyle[12pt,epsfig]{article}
\newcommand{\pphx}{p^{\uparrow}p \to hX}

\newcommand{\pup}{p^{\uparrow}}
\newcommand{\sigmaup}{\sigma^{\uparrow}}
\newcommand{\sigmadown}{\sigma^{\downarrow}}

\newcommand{\Sp}{S_{\perp}}
\newcommand{\la}{\langle}
\newcommand{\ra}{\rangle}
\newcommand{\GFt}{\widetilde{G}_F}

\newcommand{\bea}{\begin{eqnarray}}
\newcommand{\eea}{\end{eqnarray}}
\newcommand{\nn}{\nonumber}

\newcommand{\fig}[2]{\scalebox{#1}{\includegraphics{#2}}}

\newcommand{\GeV}{~{\rm GeV}}

\begin{document}
\vspace*{1.5cm}

\begin{center}

{\large \bf A phenomenological study on single transverse-spin asymmetry
\\[4mm]
for inclusive light-hadron productions at RHIC}

\vspace{1cm}

{Koichi Kanazawa$^1$ and Yuji Koike$^2$}

\vspace{1cm}

{\it $^1$ Graduate School of Science and Technology, Niigata
 University,\\
Ikarashi, Niigata 950-2181, Japan}

\vspace{0.5cm}

{\it $^2$ Department of Physics, Niigata University,
Ikarashi, Niigata 950-2181, Japan}

\vspace{2.5cm}

{\large \bf Abstract} 

\end{center}

We study the single transverse-spin asymmetry for inclusive light-hadron
productions in the proton-proton collision, $\pphx$ ($h=\pi,K,\eta$), for the
RHIC kinematics based on the twist-3 mechanism in the collinear factorization.  
The analysis includes all the contributions from the soft-gluon pole
and the soft-fermion pole for the twist-3 quark-gluon
correlation functions in the transversely polarized proton.  
After discussing the flavor decomposition and the 
$P_T$-dependence of the asymmetry obtained in the previous analysis
at the center-of-mass energy $\sqrt{S}=62.4$ and 200 GeV, 
we will give a prediction for the asymmetry at $\sqrt{S}=500$ GeV
and also for the $\eta$-meson production.
We found slightly smaller asymmetry at $\sqrt{S}=500$ GeV for $\pi^{\pm,0}$ and $K^+$
compared with those at the lower energies.  
The asymmetry for the $\eta$-meson turned out to be significantly larger than 
that for $\pi^0$.

\newpage
\section{introduction}

In the last decades, large single transverse-spin asymmetries (SSA) in
(semi-)inclusive reactions
have been receiving much attention
in high-energy spin physics (for a recent
review, see~\cite{BaroneBradamanteMartin2010}). 
Since the large SSA can not occur in the conventional framework for the high-energy process
based on the parton model and perturbative QCD, understanding the origin of
SSA provides us with a new opportunity to reveal the nucleon structure and the QCD dynamics.   
One of the most conspicuous SSA is the one
measured in the inclusive single-hadron production in the
proton-proton collision:
\bea
 \pup (p,\Sp) + p (p') \to h (P_h) + X ,
 \label{AN}
\eea
where $\Sp$ is the transverse-spin vector of the polarized proton
and $h=\pi$, $K$ and $\eta$ etc.  
The SSA for this process is characterized by 
$A_N\equiv (\sigmaup-\sigmadown)/(\sigmaup+\sigmadown) \equiv
\Delta\sigma/\sigma$, where $\sigma^\uparrow(\sigma^\downarrow)$ 
is the cross section obtained with the proton polarized along the spin vector $\Sp$ ($-\Sp$).  
The FNAL-E704 collaboration reported the first data for
$A_N$ for $\pup p\to \pi X$ and $\bar{p}^\uparrow p\to \pi X$ 
at the center-of-mass energy $\sqrt{S}=20$ GeV, which is
as large as 30 \%
in the forward direction of the polarized proton or antiproton~\cite{E7041991,E7041991charge,E7041996,E7041998}.  
RHIC at Brookhaven National Laboratory also reported a similar magnitude of $A_N$ at 
$\sqrt{S}=$62.4, 200 GeV \cite{Star2004,Phenix2005,Star2008,Brahms2008,Star2009,Phenix2010}.

In our recent paper\,\cite{KanazawaKoike2010} (hereafter referred to as KK10), we have presented a 
numerical analysis of the RHIC $A_N$ data for 
$\pup p\to h X$ ($h=\pi,\ K$) at $\sqrt{S}=62.4$ and $200$ GeV
in the framework of the collinear factorization.
Since the unpolarized cross section for this process has been well-described
by the next-to-leading-order (NLO) QCD analysis in the collinear factorization\,\cite{JagerSchaferStratmannVogelsang2003}, 
the analysis of $A_N$ in the framework can be taken as a guide to 
clarify the origin of the asymmetry.  
In the collinear factorization, SSA appears as 
a twist-3 observable and is described in terms of multiparton correlation 
functions\,\cite{EfremovTeryaev1982,QiuSterman1992,EguchiKoikeTanaka2007}.  (For the
detailed formalism of the twist-3 calculation 
demonstrating the gauge invariance and the factorization property, 
see \cite{EguchiKoikeTanaka2007,
KoikeTanaka2007Master,BeppuKoikeTanakaYoshida2010}.)
In KK10,
we focused on the contribution from the quark-gluon correlation functions
$G_F^a(x_1,x_2)$ and $\GFt^a(x_1,x_2)$ with quark-flavor $a$ ($a=u,d,s,\bar{u},\bar{d},\bar{s}$)
in the transversely polarized nucleon.  
They are defined from the light-cone correlation functions
of the form $\sim \la \bar{\psi}^a gF^{\perp +}\psi^a\ra$ in the nucleon, and 
$x_1$ and $x_2-x_1$ represent, respectively, the light-cone momentum fractions
carried by the quark and gluon lines coming out of the nucleon. 
These functions 
satisfy the symmetry property $G_F^a(x_1,x_2) =G_F^a(x_2,x_1)$
and $\GFt^a (x_1,x_2) = -\GFt^a (x_2,x_1)$.  
(For the definition and the basic properties of these twist-3 functions,
see KK10 and \cite{EguchiKoikeTanaka2007,EguchiKoikeTanaka2006}.) 
In the twist-3 mechanism for SSA, the contribution from these two functions
to the single-spin-dependent cross section $\Delta{\sigma}$ 
occurs from a pole part of an internal propagator in the hard part,
which reflects the naively $T$-odd nature of SSA.  
Such poles are classified into the soft-gluon-pole (SGP) and the 
soft-fermion-pole (SFP), which fix the momentum fractions at 
$x_1=x_2$ and $x_i=0$ ($i=$1 or 2), respectively.  
Because of the symmetry property, $\GFt^a$ does not contribute through SGP.
Therefore
the cross section can be
written as\,\cite{QiuSterman1998,KanazawaKoike2000E,KouvarisEtal2006,KoikeTanaka2007,KoikeTomita2009}
\bea
 \Delta\sigma
&=& \sum_{a,b,c} \left( G^a_{F}(x,x)-x\frac{dG^a_{F}(x,x)}{dx} \right) \otimes f^b(x')
 \otimes D^c(z) \otimes \hat{\sigma}^{\rm SGP}_{ab\to c} \nn\\
 &+& \sum_{a,b,c} \left( G^a_{F}(0,x)+\widetilde{G}^a_{F}(0,x) \right)
 \otimes f^b(x') \otimes D^c(z) \otimes
 \hat{\sigma}^{\rm SFP}_{ab\to c} , \label{formula}
\eea
where $f^b(x')$ is the twist-2 unpolarized parton distribution in the unpolarized nucleon
and $D^c(z)$ is the twist-2 unpolarized fragmentation function for the final hadron
with the parton labels $b$ and $c$ $(b,c=q,\bar{q},g)$.  
The scale dependence of each nonperturbative function is implicit.
As shown in (\ref{formula}), the SGP contribution
appears in the combination of $G_F^a(x,x)-x{d G^a_F(x,x)\over dx}$
~\cite{KouvarisEtal2006,KoikeTanaka2007}, while
the SFP contribution appears in the form $G^a_{F}(0,x)+\widetilde{G}^a_{F}(0,x)$\,\cite{KoikeTomita2009}.
In KK10, we have performed a fit of the RHIC $A_N$ data for the pion and kaon
productions based on Eq.~(\ref{formula}) and found
that the existing data have been well reproduced by the combination of the two effects.   
Although the main contribution is from the SGP contribution, the SFP contribution also plays
an important role which can not be substituted by the SGP one.  
Since we could not get a good fit by the SGP contribution only, we shall focus on 
FIT 1 of KK10 which includes both SGP and SFP contributions below.  
This analysis used the GRV98 for the unpolarized parton distribution\,\cite{GlueckReyaVogt1998} and
the DSS fragmentation function\,\cite{FlorianSassotStratmann2007PIK} for $\pi$ and $K$.

The purpose of writing this paper is twofold.
The first is
to present more details of the analysis of FIT 1 in KK10\,\cite{KanazawaKoike2010}.  
In particular, we explain the role of the SGP and SFP functions for each
quark flavor 
leading to the obtained pattern of $A_N$ for the pions and kaons.   
We also show the dependence
of $A_N$ on the final hadron's transverse momentum $P_T$ in wider $P_T$ region
and explain the origin of the characteristic $P_T$-dependence.   
The second purpose is
to give a prediction for $A_N$ for pions and kaons
at $\sqrt{S}=500$ GeV and also for the $\eta$ meson at $\sqrt{S}=200$ GeV,
which are being measured at RHIC.  
Confrontation of these predictions with the forthcoming RHIC data is critically important to
test our parametrization and to clarify the mechanism of the observed $A_N$'s.

During the present study, we found that the cross section formula
for the SGP contribution used in KK10 (Eq.(6) of \cite{KanazawaKoike2010}) has an error
in the overall sign.  (We have been using the convention for
the covariant derivative $D_\mu=\partial_\mu-igA_\mu$ and the
$\epsilon$-tensor $\epsilon_{0123}=1$, but Eq.(6) of \cite{KanazawaKoike2010} is not
consistent with this convention. ) 
Since the SGP and the SFP functions are the independent functions, 
the correction of this sign error
can be done by simply reversing the sign of the extracted SGP functions
without affecting any results for the obtained asymmetries.  
For completeness we will show in Sec.2 the corrected SGP function for each quark and antiquark flavor
together with the SFP functions.  
In the meantime, the sign error in \cite{KouvarisEtal2006} for the extracted SGP function
was also reported in a recent paper \cite{KangQiuVogelsangYuan2011}.
By our present correction, our SGP cross section formula becomes consistent with
what is claimed in \cite{KangQiuVogelsangYuan2011}.

The remainder of this paper is organized as follows:
In Sec. 2, we present the flavor decomposition of the obtained $A_N$ for the pions and kaons, and 
discuss the role of each function in the asymmetry.   
In Sec. 3, we show the $P_T$-dependence of $A_N$ for $\pi^0$ in the wider $P_T$ range and clarify the origin 
of its characteristic behavior consistent with the RHIC data\,\cite{Star2008}.  
In Sec. 4, we will give a prediction of $A_N$'s at $\sqrt{S}=500$ GeV
using the parametrization of FIT 1 of KK10,
anticipating its measurement at RHIC in the near future.
In Sec. 5, we present a prediction for $A_N$ for the $\eta$ meson at $\sqrt{S}=200$ GeV,
which reproduces the observed tendency of the preliminary RHIC data\,\cite{Star2009}.  
Section 6 is devoted to a brief summary.

\begin{figure}[!h]
\begin{center}

\fig{0.7}{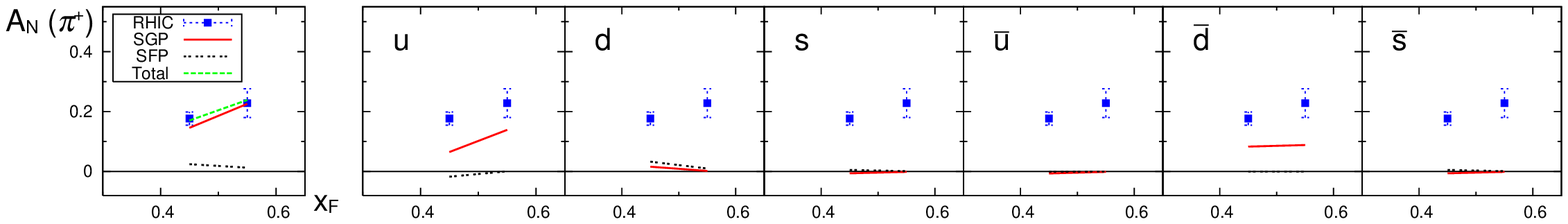}

\vspace{12pt}

\fig{0.7}{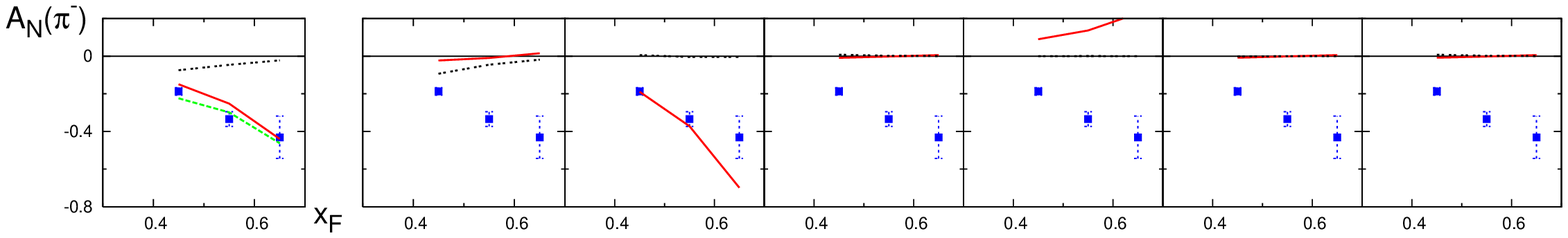}

\vspace{12pt}

\fig{0.7}{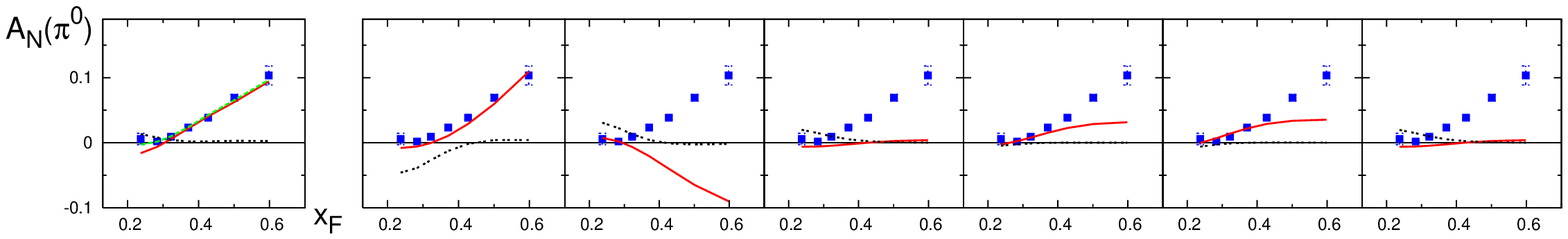}

\vspace{12pt}

\fig{0.7}{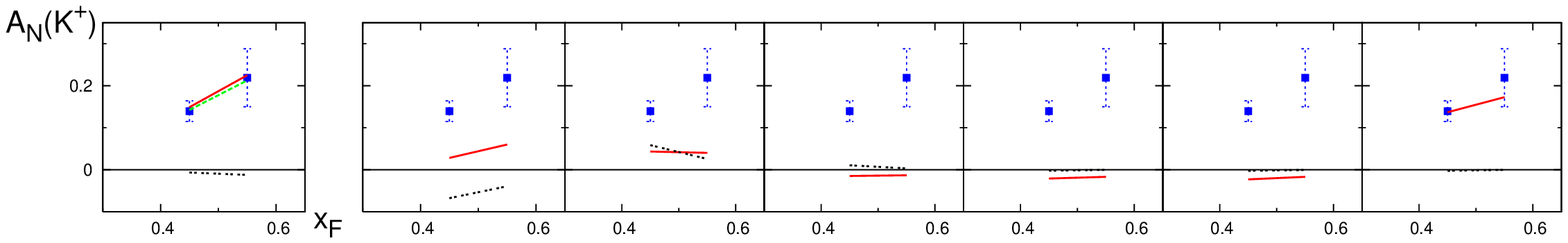}

\vspace{12pt}

\fig{0.7}{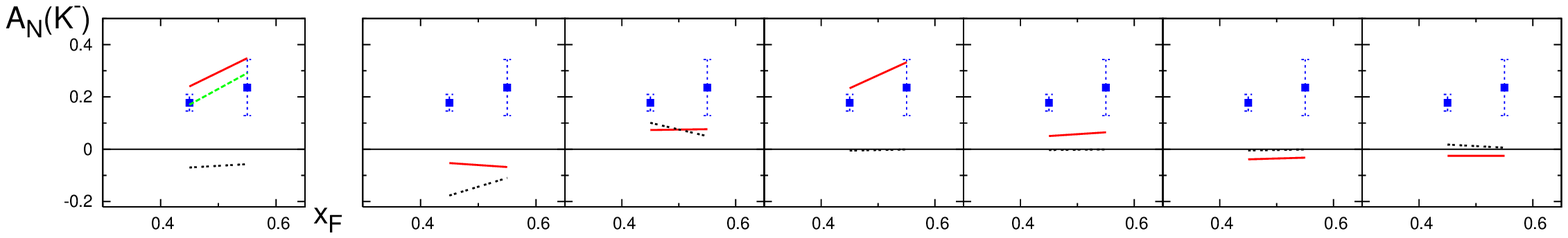}

\vspace{4pt}

\caption{Decomposition of $A_N$ at $x_F>0$ obtained in FIT 1 of KK10~\cite{KanazawaKoike2010}
into the SGP and SFP contributions (left panels)
which are decomposed further into each quark-flavor component (right panels).  
The solid and dotted lines show the
 contributions from the SGP and SFP components, 
respectively.  The dashed line in the left panels shows the total $A_N$. 
Data points from RHIC~\cite{Star2008,Brahms2008} are also shown. \label{fd}
}
\end{center}
\end{figure}

\begin{figure}
 \begin{minipage}{0.32\hsize}
  \begin{center}
   \scalebox{0.9}{\includegraphics{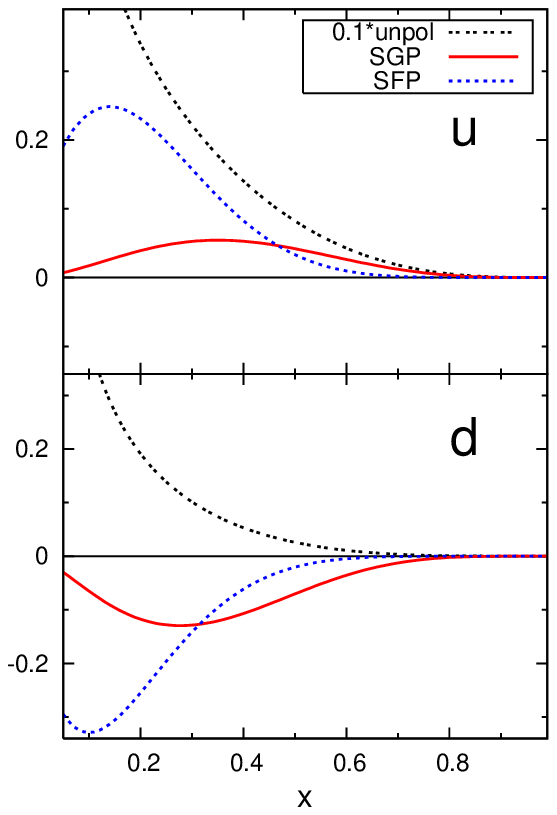}}
  \end{center}
 \end{minipage}
 \begin{minipage}{0.32\hsize}
 \begin{center}
  \scalebox{0.9}{\includegraphics{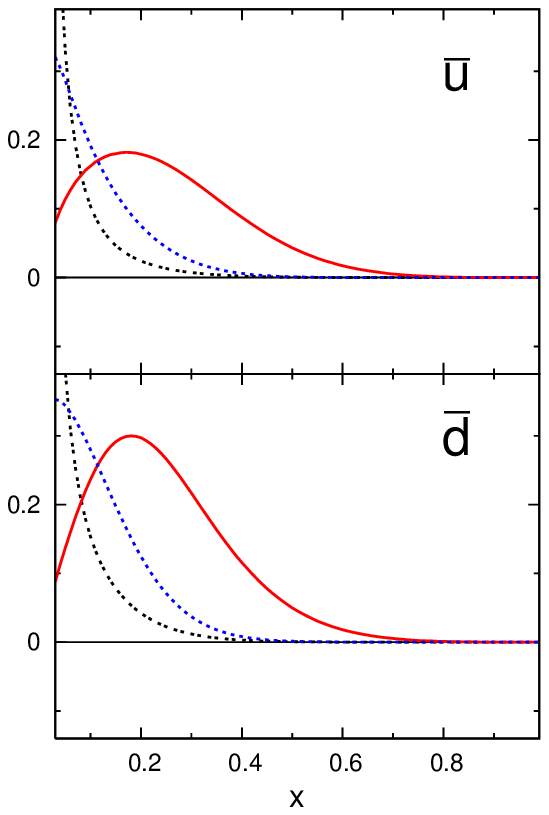}}
 \end{center}
 \end{minipage}
 \begin{minipage}{0.32\hsize}
 \begin{center}
  \scalebox{0.9}{\includegraphics{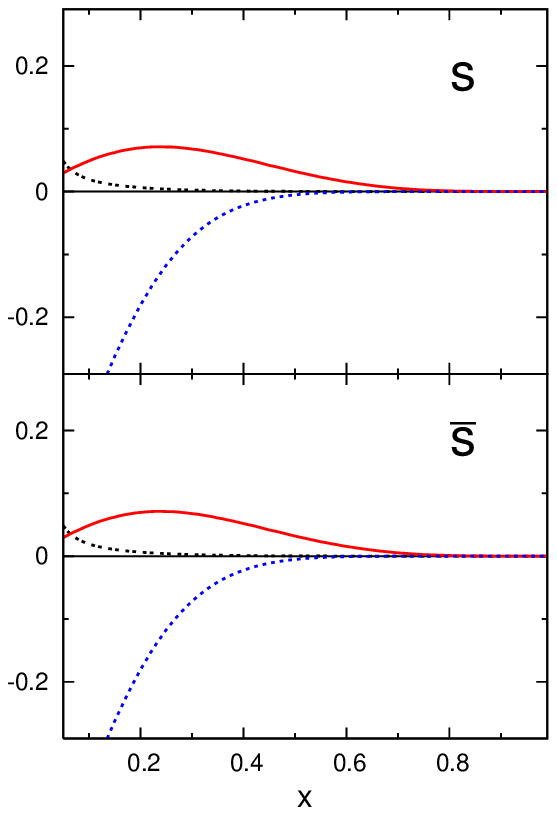}}
 \end{center}
 \end{minipage}
 \caption{The SGP function $G_{F}^{a}(x,x)$
and the SFP function $G_{F}^{a}(0,x)+\widetilde{G}_{F}^{a}(0,x)$
at the scale $\mu^2=2.4$ {GeV}$^2$
for each quark flavor obtained in FIT 1 of KK10 in comparison to
the unpolarized quark distribution $f_a(x)$ [scaled by factor 1/10].
\label{func}}
\end{figure}

\section{Flavor structure of the asymmetry}
We begin our discussion by 
the flavor structure of the asymmetry for $\pi^{\pm,0}$ and $K^\pm$.  
In Fig.~\ref{fd}, we show the decomposition of 
$A_N$ into the SGP and SFP contributions for each quark and antiquark flavor.  
The left panels show the decomposition of $A_N$
into the total SGP and SFP contributions,
while the right panels show the contribution from each quark and antiquark flavor
both for the SGP and SFP contributions.
To understand the pattern of the flavor structure, we show the
SGP and SFP functions obtained in FIT 1 of KK10 for each quark and antiquark flavor in Fig. 2,
by correcting the sign error mentioned at the end of the introduction.       
First, as seen in the top two panels of Fig.~\ref{fd}, the major contributions for $\pi^\pm$
come from the SGP components for the favored quark flavors:
$A_N$ for $\pi^+$ is dominated by the SGP contribution for the $u$ and $\bar{d}$ quarks, 
and $A_N$ for ${\pi^-}$ is dominated by that for the $d$ and $\bar{u}$ quarks.  
Note here the significance of the $\bar{u}$ and $\bar{d}$ components, which results from
the large SGP functions for these ``sea" flavors in the nucleon at large $x$
as shown in Fig.~\ref{func}.    
The net SFP effect turned out to be small for $\pi^\pm$
in the region of $x_F>0.4$ for which the RHIC data are available.  
If one looks into more details of the SFP contribution, the SFP effect brings
some net effect for $\pi^-$, while it is much smaller for $\pi^+$.  This feature of the SFP contribution
can also be seen for $K^\pm$.  (See discussions below.)
The above feature of the SGP dominance can also be seen in $A_N$ for $\pi^0$:   
All the SGP contributions from $u$, $d$, $\bar{u}$ and $\bar{d}$ bring large contributions to
$A_N^{\pi^0}$ in the large $x_F$ region.  
The net SFP effect can be
found in the small $x_F$ region for $\pi^0$: 
It brings some effect at $x_F<0.4$ which cancels the SGP contribution, 
making the $A_N$ small.
This feature can be understood if one remembers
that the partonic hard cross section for SFP is much larger than that for
SGP in many relevant channels, although the derivative of the SGP function enhances 
its contribution.

The twist-3 functions for the sea quark flavors in the nucleon become more important for the kaons.
In particular, the observed large $A_N$ for $K^-$ should come from the
sea quarks due to its flavor structure, $K^-\sim \bar{u}s$.
As seen in Fig.~\ref{fd}, 
the largest contribution to $A_N$ for $K^+$ and $K^-$ comes from the SGP functions
for the $\bar{s}$ and $s$ quarks, respectively, and for $K^+$ even the SGP contribution
from the $u$-quark is significantly smaller than that from the $\bar{s}$-quark.  
This is due to the much larger strangeness component for the kaon in the DSS fragmentation
function as well as the sizable magnitude of the SGP functions for the $s$ and $\bar{s}$ quarks in the 
nucleon.   Combined with the result for the pions, 
one may say that $A_N$ for the pions and kaons reflects the
flavor structure of the fragmentation functions rather than that of the 
SGP functions.  
This is in contrast to the unpolarized cross section where the contribution from the
valence quark flavors in the nucleon
dominate in the forward rapidity region. 

\begin{figure}[!t]
 \begin{center}
  \fig{1.0}{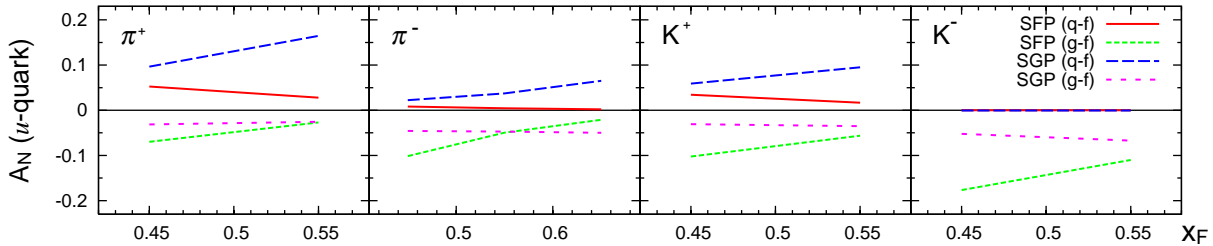}
  \caption{Decomposition of the $u$-quark SGP and SFP contribution to $A_N$ 
  at $\sqrt{S}=62.4$ GeV shown in Fig.~\ref{fd}
into the quark and gluon-fragmentation channel. 
\label{udeco}}
 \end{center}
\end{figure}

The contribution of the SFP effect shows a very different pattern from
the SGP contribution.  As shown in Fig. 1, the net SFP effect survives for $K^-$ and $\pi^-$
at $x_F>0.4$, while it is negligible for $K^+$ and $\pi^+$.
This net effect for the former is caused by the SFP function for the $u$-quark flavor 
through the gluon-fragmentation channel in combination with 
the large gluon component in the DSS fragmentation function, in particular, for the kaon.  
(The SFP function for the $d$-quark is also large and partly cancels the $u$-quark contribution. 
However, the $u$-quark function has a longer range than the $d$-quark function, and thus the former effect
is larger.) 
As shown in \cite{KoikeTomita2009}, the SFP partonic 
hard cross section in the gluon-fragmentation
channel is extremely large and has the opposite sign compared with that in the quark
fragmentation channel.  For $K^+$ and $\pi^+$, the $u$-quark SFP contribution
in the gluon-fragmentation channel is mostly canceled by that in the $u$-quark-fragmentation
channel for which $K^+$ and $\pi^+$ have a large fragmentation function as a favored flavor, 
while the former contribution is not canceled by the latter for $K^-$ and $\pi^-$.  
To see this feature, we showed in Fig. \ref{udeco} the decomposition
of the SGP and SFP contribution for the $u$-quark into the quark-fragmentation channel and
the gluon-fragmentation channel.  As seen from this figure, the $u$-quark SFP contribution
via the gluon-fragmentation channel survive for $K^-$ and $\pi^-$,
while they are canceled by that via the quark-fragmentation channel for $K^+$ and $\pi^+$.

Summarizing the flavor structure of $A_N$, 
the SGP contribution for the favored quark flavors is the dominant one, 
while the SFP contribution for the valence $u$ and $d$ quarks is also
significant via the gluon-fragmentation channel.  This SFP effect
survive for $K^-$ and $\pi^-$ but is canceled by that in the quark-fragmentation channel
for $K^+$ and $\pi^+$.

\section{$P_T$-dependence} 

\begin{figure}[!t]
 \begin{center}
  \fig{1.0}{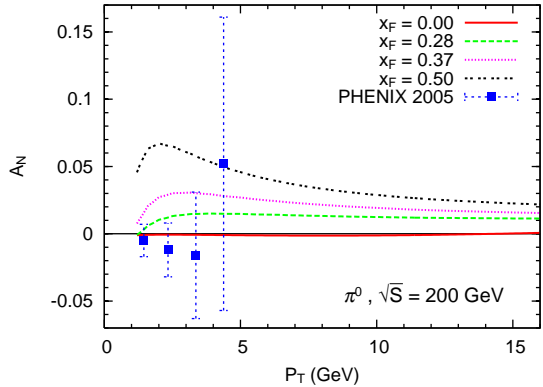}
  \caption{$P_T$-dependence of $A_N^{\pi^0}$ at
  $\sqrt{S}=200\GeV$ and four fixed $x_F$ 
together with the PHENIX data at $x_F=0$~\cite{Phenix2005}. \label{pt}}
 \end{center}
\end{figure}

The $P_T$-dependence of $A_N$ provides an important test for the mechanism of the asymmetry. 
In Ref.~\cite{KanazawaKoike2010}, we have found that our predictions for the $\pi^0$
production shows a good
agreement with the RHIC-STAR data at $\sqrt{S}=200$ GeV. 
In Fig.~{\ref{pt}}, we show the $P_T$-dependence of $A_N$ for $\pi^0$
in the wider range of $P_T$ at $x_F=0$, 0.28, 0.37 and 0.50 together with
the RHIC-PHENIX data at
$x_F=0$~\cite{Phenix2005}.
As seen in the figure, calculated $A_N$ at $x_F=0$ stays zero in the whole $P_T$ region,  
which is in agreement with the PHENIX data.

A conspicuous feature of the $P_T$-dependence is that the calculated 
$A_N$ once increases up to $P_T\sim$ a few GeV and then decreases slowly, which is
in agreement with the STAR data.  
This behavior is in conflict with our
naive expectation that $A_N$ should behave like
$1/P_T$ as a function of $P_T$ by reflecting its twist-3 nature.
We have found that the contribution from the gluon-fragmentation
channels plays an important role to
cause this peculiar $P_T$-dependence of $A_N$.   
It brings a negative contribution to the positive polarized cross section
$\Delta\sigma$ which is the numerator of $A_N$, while it brings a positive one to the
unpolarized cross section $\sigma$.  
With the de Florian-Sassot-Stratmann (DSS) fragmentation function used which has a large gluon component, 
their contributions decrease very rapidly as the scale $\mu=P_T$ becomes higher.  
Correspondingly, $\Delta\sigma$ decreases relatively slowly 
around $P_T\simeq 1-3$ GeV 
compared with the denominator, 
even though the twist-3 cross section accompanies the factor $1/P_T$.  
Such behavior is absent in the previous study by
Kouvaris~{\it et al}~\cite{KouvarisEtal2006} which used 
Kretzer's fragmentation function~\cite{Kretzer2000} whose gluon
fragmentation function is small. 
For comparison we calculated $A_N$ by fixing the scale of the 
distribution and fragmentation functions 
at $\mu=1$ GeV (instead of $\mu=P_T$) and found that
$A_N$ decreases monotonously as $P_T$ increases and diverges toward $P_T\to 0$.

Another interesting point in Fig.~\ref{pt}
is that $A_N$ does not decrease as fast as $1/P_T$ in the high $P_T$ region.  
This is because of the coexistence of the
two effects proportional to $M_NP_T/(-T)$
and $M_NP_T/(-U)$ in the twist-3 asymmetry
where $T=(p-P_h)^2$ and $U=(p'-P_h)^2$ are the Mandelstam variables:
While the former effect is much larger at small $P_T$, the latter also brings significant contribution
at large $P_T$.

\section{$A_N$ at $\sqrt{S}=500$ GeV}
Since the measurement of $A_N$ is being extended to $\sqrt{S}=500$ GeV at RHIC, 
we present a prediction for $A_N$ at this energy, using the 
SGP and SFP functions obtained from the data at $\sqrt{S}=200$ and 62.4 GeV.  

\begin{figure}[!t]
 \begin{center}
  \fig{1.0}{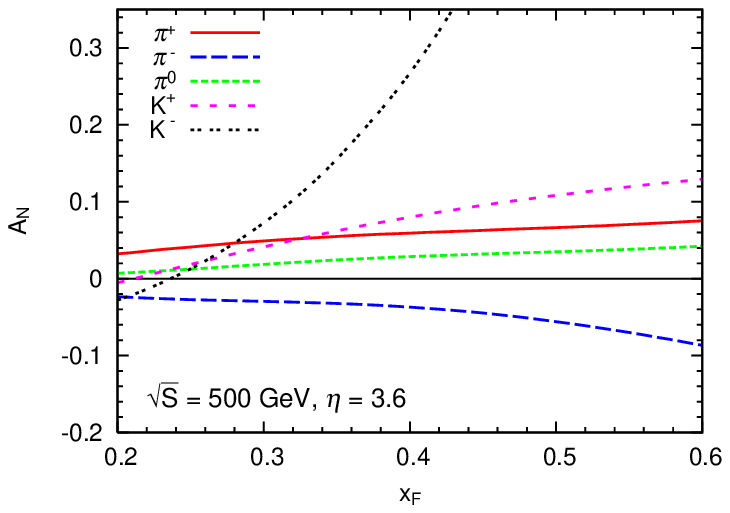}
  \fig{1.0}{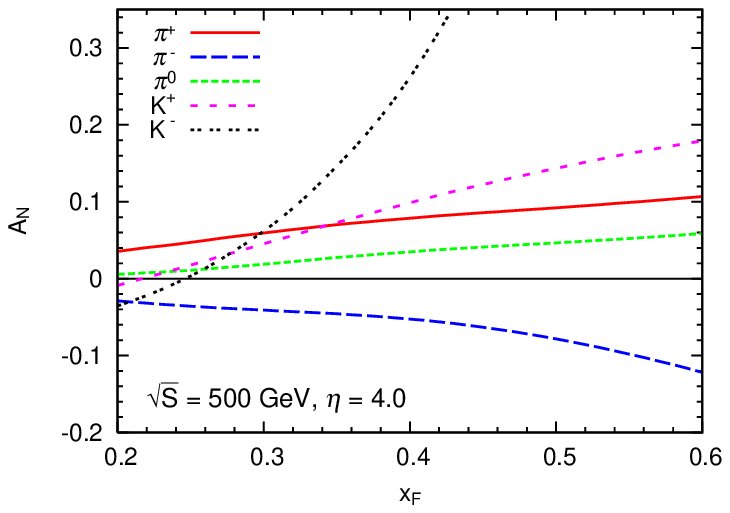} 
\end{center}
 \caption{$A_N$ for pions and kaons at $\sqrt{S}
 = 500\GeV$ and the pseudorapidity
 $\eta = 3.6$ (left) and $\eta=4.0$ (right). 
\label{AN500}}
\end{figure}

In Fig.~\ref{AN500}, we show the result of $A_N$ for pions and kaons 
as a function of $x_F$ for the pseudorapidity $\eta=3.6$ and $\eta=4.0$.  
For the pions, the general trend of the asymmetry is the same as those at lower energies,
while the magnitude of the asymmetry becomes smaller 
at each $x_F$.  This is because a larger value of $\sqrt{S}$
corresponds to larger $P_T$ for a fixed $x_F$ and thus the polarized cross section is more power 
suppressed at higher $\sqrt{S}$.

$A_N$ for kaons turned out to become larger than those for charged pions 
in the large $x_F$ region, in particular, for the $K^-$ meson.  
This larger $A_N$ for $K^\pm$ is mainly due to the faster
decrease of the unpolarized cross section for the $K^\pm$ production 
compared with the $\pi^\pm$ production with increasing $\sqrt{S}$ from 62.4 to 500 GeV.  
This decrease of the unpolarized cross section 
becomes even more manifest for the $K^-$ production, 
since the unpolarized cross section in the forward region is dominated by the contribution
from the valence flavors in the nucleon and the sea contribution
is more suppressed at higher scale $\mu=P_T$ at large $x_F$.  
Therefore $A_N$ for $K^-$ becomes much larger than that for $K^+$.
We found that $A_N$ for $K^-$ overshoots one in the large $x_F$ region.  
In our present analysis, the constraint on the SGP and SFP functions
for the $s$ and $\bar{s}$ quarks comes only from a very limited number of the
kaon data with the assumption that they are the same for the $s$ and $\bar{s}$ quarks.  
With a larger number of data points, this unfavorable feature should be corrected.
Nevertheless, we expect the relation 
$A_N^{K^-}> A_N^{K^+} \geq A_N^{\pi^+}$ at $\sqrt{S}=500$ GeV for the reason stated above.

\begin{figure}[!t]
 \begin{center}
  \fig{1.0}{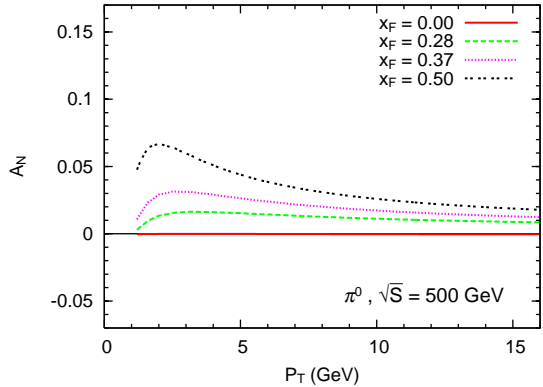}
 \end{center}
 \caption{$P_T$-dependence of $A_N^{\pi^0}$ for $\sqrt{S}
 = 500$ GeV at $x_F=$0, 0.28, 0.37 and 0.5. \label{pt500}}
\end{figure}

Finally we show in Fig.~\ref{pt500} the $P_T$-dependence of $A_N$ for $\pi^0$ at $\sqrt{S}=500$ GeV.
As in the case of $\sqrt{S}=200$ GeV, $A_N$ once increases toward $P_T\simeq$ a few GeV and then decreases
slowly.

\section{$A_N$ for $\eta$ meson}

The first data of $A_N$ for the $\eta$-meson production was 
reported by the FNAL-E704 collaboration for the 
$pp$ and $\bar{p}p$ collisions~\cite{E7041998}. They found a
clear indication that $A_N^\eta$ for the $pp$ collision
is positive and substantially larger than $A_N^{\pi^0}$.  
Recently
the RHIC-STAR collaboration reported a preliminary data 
for $A_N^\eta$ at $\sqrt{S}=200$ GeV, which also shows $A_N^\eta$ is larger than 
$A_N^{\pi^0}$ in the large $x_F$ region by about factor two or more\,\cite{Star2009}.
Here we present an estimate for $A_N^\eta$ based on Eq.~(\ref{formula}) using 
the SGP and SFP functions obtained in KK10. 
In this scenario, the difference between $A_N^\eta$ and $A_N^{\pi^0}$
comes from the difference between the difference in the fragmentation functions.
As the fragmentation function for the $\eta$-meson, 
we adopt the recent Aidala- Ellinghaus- Sassot- Seele- Stratmann parametrization~\cite{AidalaEtal2011}
which is obtained by
NLO QCD analysis of the existing data, 
assuming the equal contributions from all quark and antiquark flavors as
$D^{\eta}_u=D^{\eta}_{\bar{u}}=D^{\eta}_d=D^{\eta}_{\bar{d}}=D^{\eta}_s=D^{\eta}_{\bar{s}}$.  
Although we used the leading-order (LO) formula for the
twist-3 cross section (\ref{formula}), we use the above only available NLO parametrization
for the $\eta$-fragmentation function in this first study of $A_N^\eta$.
\footnote{Since the LO and NLO parametrizations for $\pi$ in \cite{FlorianSassotStratmann2007PIK} 
are not very different from
each other, we expect the use of the NLO fragmentation function for $\eta$
would not cause much error.}

\begin{figure}[!t]
 \begin{center}
  \fig{1.0}{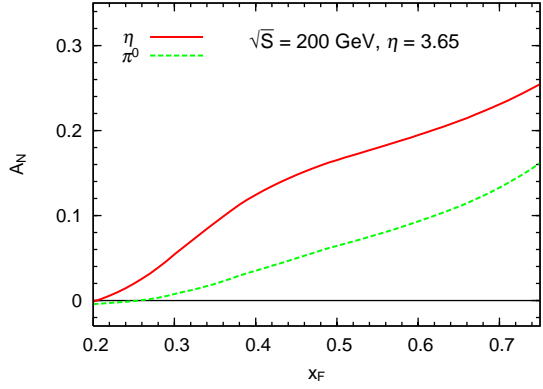}
 \end{center}
 \caption{$A_N$ for the $\eta$-meson at
 $\sqrt{S}=200\GeV$ and the pseudorapidity $\eta=3.65$ in comparison with $A_N$ for $\pi^0$. 
\label{ANeta}}
\end{figure}

\begin{figure}[!t]
 \begin{center}
  \fig{1.0}{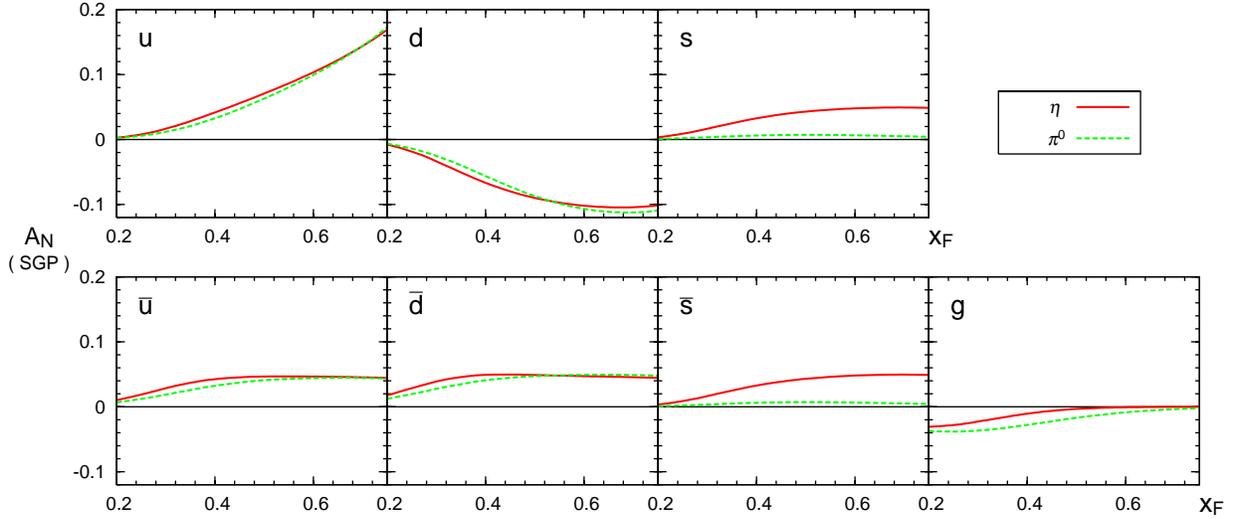}
 \end{center}
 \caption{Decomposition of $A_N^{\eta}$ and $A_N^{\pi^0}$ 
into each fragmentation channel at
$\sqrt{S}=200\GeV$ and the pseudorapidity $\eta=3.65$.
\label{fc}}
\end{figure}

{}Figure~\ref{ANeta} shows the calculated $A_N^\eta$ as a function of $x_F$ 
in comparison with $A_N^{\pi^0}$ at
$\sqrt{S}=200\GeV$ and $\eta=3.65$.  In the large
$x_F$ region, $A_N^{\eta}$ is approximately 1.5 times larger than
$A_N^{\pi^0}$.  This is consistent with the observed tendency of $A_N^\eta$ at RHIC, 
although its magnitude still seems insufficient
to explain the RHIC-STAR data\,\cite{Star2009}. 
As in the case of $A_N^{\pi^0}$, the SGP contribution dominates $A_N^\eta$.
To see the origin of the difference between $A_N^{\eta}$ and $A_N^{\pi^0}$, 
we have shown in Fig.~\ref{fc} the decomposition of the SGP contribution to $A_N^\eta$ and $A_N^{\pi^0}$
into the contribution from each fragmentation channel.  From this figure, 
one sees that the difference between the two mesons comes from 
the large contribution in the strangeness fragmentation channel for the $\eta$-meson.  
This large strangeness contribution 
is caused by the combination of the large strangeness component in the
$\eta$-meson fragmentation function and the
large SGP functions for $s$ and $\bar{s}$ which were responsible for $A_N^{K^\pm}$.  
Therefore in the present analysis,
there is a strong correlation between the large $A_N$ for the kaon and the $\eta$-meson.  
As another possible origin of the difference between $A_N^\eta$ and $A_N^{\pi^0}$,
the twist-3 fragmentation function may contribute\,\cite{KangYuanZhou2010}, 
for which the complete cross section formula is not available yet. 
This effect may explain the remaining difference between the two asymmetries.  
For the complete clarification of the asymmetries, one needs to
perform a global analysis of 
a greater variety of data using the complete cross section formula, 
which is beyond the scope of this study.

\section{Summary}

In this paper, we have studied the SSA for the inclusive pion, kaon and
$\eta$-meson productions in the $pp$ collision at typical kinematical regime of
RHIC, using the SGP and SFP functions for the twist-3 quark-gluon correlation functions
obtained in our previous analysis.  
We first discussed the role of the SGP and SFP functions for each
quark and anti-quark flavor.  It turned out that
the SGP contributions from both ``valence" and sea flavors
in the nucleon play an important role to reproduce the observed pattern of the asymmetries,
and that the main contribution to $A_N$ for each meson reflects
more the flavor content in the DSS fragmentation function.   
Our analysis reproduces the observed $P_T$-dependence of RHIC $A_N$ data for $\pi^0$.  
We found that the large contribution from the gluon-fragmentation channel
both in the unpolarized and polarized cross sections plays a crucial role
to cause the peculiar behavior of $A_N$ around $P_T\simeq$ a few GeV.  

We also presented a prediction of $A_N$ for the kaons and pions at $\sqrt{S}=500$ GeV,
which is being measured at RHIC.  We found a slightly smaller asymmetry for the pion 
compared with those at $\sqrt{S}=62.4$ and 200 GeV.  
For the kaons, $A_N$ was found to become larger than those for the pions
in the large $x_F$-region at $\sqrt{S}=500$ GeV, especially $A_N$ for $K^-$ becomes much larger
because of the small unpolarized cross section.  
$A_N$ for the $\eta$-meson turned out to be larger than that for $\pi^0$
by about factor 1.5, which partially explains the tendency of the RHIC data. 
This larger $A_N$ originates from the large strangeness fragmentation function for the $\eta$ meson
combined with the sizable strangeness SGP function in the polarized nucleon.  

Present analysis is based on the assumption that
the whole $A_N$ for $p^\uparrow p\to hX$ ($h=\pi$, $K$ and $\eta$)
originates from the twist-3 quark-gluon correlation functions
in the polarized nucleon.  To determine these functions, the SSA data on the Drell-Yan
process and the direct-photon production in the $pp$-collision is
extremely useful, since there is no fragmentation ambiguity 
associated with the final hadron in these processes. 
For the complete understanding
of SSAs, a global analysis of variety of processes based on the complete formula
is needed.


\section*{Acknowledgements}

We thank C.~A.~Aidala and M.~Stratmann for their correspondence and
providing their fortran code of the fragmentation function for $\eta$-meson.
The work of Y.~K. is supported in part by the Grant-in-Aid for
Scientific Research No.~23540292 from the Japan Society for the
Promotion of Science.


\end{document}